\documentclass[a4paper,fleqn,usenatbib]{mnras}

\usepackage{newtxtext,newtxmath}
\usepackage{mathptmx}

\usepackage[T1]{fontenc}
\usepackage{ae,aecompl}

\usepackage{graphicx}	
\usepackage{amsmath}	
\usepackage{amssymb}	

\title[Column Density PDF]{The Effect of Magnetic Fields and Ambipolar Diffusion on the Column Density Probability Distribution Function in Molecular Clouds }
\author[S . Auddy  et al.]{
Sayantan Auddy,$^{1,2}$\thanks{E-mail: sauddy3@uwo.ca}
Shantanu Basu,$^{1}$\thanks{E-mail: basu@uwo.ca}
Takahiro Kudoh$^{3}$\thanks{Email: kudoh@nagasaki-u.ac.jp}
\\
$^{1}$Department of Physics and Astronomy,The University of Western Ontario, London, ON N6A 3K7, Canada\\
$^{2}$Harvard-Smithsonian Center for Astrophysics, 60 Garden Street, Cambridge, MA 02138, USA\\
$^{3}$Faculty of Education, Nagasaki University,1-14 Bunkya-manchi, Nagasaki 852-8521, Japan\\
}

\date{Accepted XXX. Received YYY; in original form ZZZ}

\pubyear{2016}

\begin{document}
\label{firstpage}
\pagerange{\pageref{firstpage}--\pageref{lastpage}}
\maketitle

\begin{abstract}
	 Simulations generally show that non-self-gravitating clouds have a lognormal column density ($\Sigma$) probability distribution function (PDF), while self-gravitating clouds with active star formation develop a distinct power-law tail at high column density.
	Although the growth of the power law can be attributed to gravitational contraction leading to the formation of condensed cores, it is often debated if an observed lognormal shape is a direct consequence of supersonic turbulence alone, or even if it is really observed in molecular clouds.
	In this paper we run three-dimensional magnetohydrodynamic simulations including ambipolar diffusion with different initial conditions to see the effect of strong magnetic fields and nonlinear initial velocity perturbations on the evolution of the column density PDFs. Our simulations show that column density PDFs of clouds with supercritical mass-to-flux ratio, with either linear perturbations or nonlinear turbulence, quickly develop a power-law tail such that $dN$/$d \log \Sigma \propto \Sigma^{- \alpha}$ with index $\alpha \simeq 2$. Interestingly, clouds with subcritical mass-to-flux ratio also proceed directly to a power-law PDF, but with a much steeper index $\alpha \simeq 4$. This is a result of gravitationally-driven ambipolar diffusion. However, for nonlinear perturbations with a turbulent spectrum ($v_{k}^{2} \propto k^{-4}$), the column density PDFs of subcritical clouds do retain a lognormal shape for a major part of the cloud evolution, and only develop a distinct power-law tail with index $\alpha \simeq 2$ at greater column density when supercritical pockets are formed.
 
\end{abstract}

\begin{keywords}
	ISM: clouds -- ISM: magnetic fields -- magnetohydrodynamics (MHD) -- stars: formation
\end{keywords}



\section{Introduction}

In recent years there has been a growing interest in the study of the column density probability distribution function (PDF) of molecular clouds. 
This function is significant in current theories of star formation as it is used to explain the initial mass function \citep{pad02,hen08}, star formation rates \citep{pad11,hen11,fed12} and star formation efficiencies \citep{fed13} of molecular clouds.  The PDF is the normalized histogram of the column density obtained from the 
measurement within some area of the sky that contains a molecular cloud. 
One way to measure the column density of a molecular cloud is using dust extinction and the reddening of the light of background stars in the near IR \citep{alv01,kai09,alv14}. Furthermore, mid-infrared (IR) absorption \citep{bac00}, millimetre continuum emission \citep{war99} and flux measurements in optically thin lines \citep{taf02} are the other commonly used methods to measure column density. 

Early numerical investigation \citep{sem94,pad97,pas98,pad99} through non-self-gravitating purely hydrodynamic isothermal simulations showed that a lognormal density PDF is a preferred outcome of the development of hierarchical structures. The next generation of magnetohydrodynamic (MHD) simulations of the interstellar medium including self-gravity \citep{sca98,fed08} found evidence of a growing power law at high densities. It was natural to interpret that the volume density PDF follows an underlying lognormal distribution, with a departure to a power law at higher density as it develops gravitationally collapsed objects.

Recent observations by \cite{kai09} are broadly consistent with the idea that the column density PDF has an underlying lognormal shape with an additional power law at high column density. 
They identified that active star-forming clouds have an excess of high column densities, which manifests in the nonlognormal wings of the PDF. In contrast, quiescent clouds without active star formation are fit well by a lognormal distribution over the whole range of observed column density. A \textit{Herschel-SPIRE} survey of the Mon R2 giant molecular cloud \citep{pok16} found that the gas column density PDF is lognormal, but with a power-law tail with best-fit index $ \alpha = 2.15$ above $\sim 10^{21}\, \rm{cm}^{-2}$. These observations are consistent with the evolutionary trend where turbulent motions play the main role in shaping the cloud in the early stages, but core formation is dominated by gravity and possibly magnetic fields. 
Several numerical studies \citep{tas10,par11,war14} have subsequently shown that a power-law tail develops over time and its strength grows as the rate of star formation activity increases. 

A recent survey by \cite{lom15} reconfirms the results from \cite{kai09} that at a high extinction the PDFs are best fit with a power law. They measure the column density in terms of the K-band extinction $A_K$
from the dust emission maps of Herschel and Planck data, and show that for $A_K \geqslant 0.2$ mag, the PDFs ($d N$/$d \log A_K \propto A_{K}^{-\alpha}$) have power-law indices with $\alpha \approx 2$, but
clouds with lower star formation activity, i.e. Polaris and Pipe, have $\alpha = 3.9$ and $\alpha=3.0$, respectively.  \citet{kon15} also find a power-law PDF for the Aquila star-forming cloud using Herschel data, with index $\alpha \approx 2$. \cite{alv17} extend the idea that PDFs of molecular clouds are only a power law, with slope varying from $\alpha \approx 4$ for diffuse clouds to $\alpha \approx 2$ for clouds with active star formation. This is consistent with the fact that steeper slopes mean a lack of high density material and thereby less star formation.  
However, a physical explanation of a steeper slope in such clouds has been lacking.

A key question is whether clouds with a strong magnetic field exhibit distinct features of column density PDFs in comparison to clouds with a weak magnetic field. The lognormal feature is often interpreted as a direct imprint of supersonic turbulence, which is believed to dominate the evolution of observed clouds \citep{sem94}. However, the recent work by \cite{tas10} points out that lognormal column density PDFs may be a more generic feature of molecular clouds and should not be interpreted as a result of supersonic turbulence alone.
Observationally there is also the claim by \citet{alv17} that the lognormal peak may be an artifact arising due data incompleteness, and thereby not a result of supersonic turbulence. \cite{tas10} also show that gravitationally-driven ambipolar diffusion plays a significant role in shaping the PDFs. Furthermore, a thermally bi-stable numerical simulation by \cite{par11} reveals that global gravitational contraction enhances the initial density fluctuations and results in a wider lognormal PDF and a power-law tail at later times. 
It is likely then that the column density PDFs of molecular clouds arise from a variety of initial conditions and can represent different evolutionary stages. 
Here, we explore the process of molecular cloud fragmentation based on the interplay of turbulence, gravity, and magnetic fields.  While large scale turbulence sweeps up the interstellar medium and compresses the gas into filaments and shocks, magnetic fields can provide a global support against the collapse until gravitationally-driven ambipolar diffusion leads to a runaway collapse of the densest regions of the cloud \citep[e.g.][]{nak05,kud11}. In this paper, we explore the effect of the magnetic field, gravity, and ambipolar diffusion in determining the column density PDFs. Our study follows the previous ones by \cite{kud08,kud11}. We carry out a parameter study by running a number of simulations with different initial conditions. Our main objective is to see the effect of large-scale magnetic fields and nonlinear initial perturbations on the time evolution of the column density PDF. We further investigate the differences in steepness of the power-law index of the column density PDF and connect them to different initial conditions. 
Since direct magnetic field measurements using the Zeeman effect \citep{cru12} are rarely successful, a key goal is to find a link between the structural properties of molecular clouds and the ambient magnetic field strength (or mass-to-flux ratio).
 
Our paper is organised in the following manner. The numerical model and some background theory is discussed in Section \ref{numericalModel}, and the results from the simulations are given in Section \ref{results}. We provide more discussion of the results in Section \ref{discussions} and give a summary in Section \ref{conclusion}.
\section{Theory and Numerical Model}\label{numericalModel}
Magnetic fields and ambipolar diffusion play an important role in the star formation process. They can regulate the cloud collapse and fragmentation process, control angular momentum evolution through magnetic braking, and possibly moderate the mass reservoir for stars by limiting the mass accretion from the magnetic envelope. The relative strength of gravity and the magnetic field is measured by the mass-to-flux ratio $M$/$\Phi$. There exists a critical mass-to-flux ratio ($M$/$\Phi$)$_{\rm crit}$ \citep{mes56,mou76,str66,tom88}. For $M$/$\Phi >$($M$/$\Phi$)$_{\rm crit}$, the cloud is supercritical and is prone to indefinite collapse. However, for $M$/$\Phi <$($M$/$\Phi$)$_{\rm crit}$ the cloud is subcritical and cannot collapse as long as magnetic flux freezing applies. 
For example, $M$/$\Phi <$($M$/$\Phi$)$_{\rm crit}$$= $($2 \pi G^{\frac{1}{2}}$)$^{-1}$ is required for stability against fragmentation for an infinite uniform layer that is flattened along the direction of the background magnetic field \citep{nak78}.
However, in nonideal MHD, neutral-ion slip leads to gravitationally-driven fragmentation on the ambipolar-diffusion timescale \citep{lan78,cio06,mou11}. 
In this project we focus on the fragmentation and core formation in nonideal MHD clouds. \cite{kud07} performed a three-dimensional simulation of the fragmentation and core formation in the subcritical clouds with ambipolar diffusion and gravitational stratification along the magnetic fields. \cite{kud08,kud11} did a further parameter study to demonstrate that core formation occurs faster as the strength of the initial flow speed in the cloud increases. 

\subsection{Setup for numerical simulation}
The numerical model used in this paper is similar to previous ones \citep{kud07,kud08,kud11}. We solve the three-dimensional nonideal magnetohydrodynamic (MHD) equations including self-gravity and ambipolar diffusion: 
\begin{eqnarray}
\frac{\partial \rho}{\partial t} + \textbf{v} \cdot\nabla \rho = -\rho \nabla\cdot \textbf{v} \label{continuity eqn},\\
\frac{\partial\textbf{ v}}{\partial t} + \left(\textbf{v}\cdot\nabla\right) \textbf{v} = - \frac{1}{\rho} \nabla p+\frac{1}{c\rho} \textbf{j} \times \textbf{B} - \nabla \psi \label{momemtum eqn},\\
\frac{\partial \textbf{B}}{\partial t} = \nabla \times \left(\textbf{v} \times \textbf{B}\right)+ \nabla \times \left[ \frac{\tau_{ni}}{c \rho} \left(\textbf{j} \times \textbf{B}\right)\times \textbf{B} \right]\label {f eqn}\label{magnetic ind eqn},\\
\textbf{j} = \frac{c}{4 \pi}\nabla \times \textbf{B} \label{magnetic current eqn}, \\
\nabla^{2}\psi = 4 \pi G \rho \label{poisson eqn} ,\\
p= c_s^2 \rho ,\label{eqn of state}
\end{eqnarray}
where $\rho $ is the density of the neutral gas, $p$ is the pressure, $\textbf{v}$ is the velocity, $\textbf{B}$ is the magnetic field, $\textbf{j}$ is the electric current density, $\psi$ is the self-gravitational potential and $c_s$
is the sound speed. 
Equations (\ref{continuity eqn}) and (\ref{momemtum eqn}) are the mass continuity and the momentum equations, respectively. Equation (\ref{magnetic ind eqn}) is the magnetic induction equation. 
The neutral-ion collision time in equation (\ref{magnetic ind eqn}) is given by 
\begin{equation}
\tau_{ni}=1.4 \frac{m_i+m_n}{\rho_i \langle \sigma w \rangle_{in}}
\end{equation} 
\citep[e.g.][]{bas94}, where $\rho_i$ is the ion density and $\langle \sigma w \rangle_{in} $ is the average collision rate between the ions of mass $m_i$ and neutrals of mass $m_n$. 
Furthermore, it is assumed that the temperature of the gas makes a step-like transition from a cool molecular gas to a warm surrounding medium at a height of $z_c = 2H_0$ (see Eq. [18] of \cite{kud11}) and that in the subsequent evolution each Lagrangian fluid particle is in isothermal equilibrium \citep{kud03,kud06} so that
\begin{equation}
\frac{dc_s}{dt}=\frac{\partial c_s}{\partial t} + \textbf{v}\cdot \nabla c_s=0.
\end{equation}
This means that each parcel of the molecular cloud and the surrounding warm gas retain their initial temperature. 
As an initial condition we set up the simulation box with a preferred direction of the uniform magnetic field. The self-gravitating cloud is in hydrostatic equilibrium along the direction of the magnetic field and forms a sheet-like geometry. 
The one-dimensional hydrostatic equilibrium can be calculated using the following equations:
\begin{equation}
\frac{dp}{dz}=\rho g_z,\ \frac{dg_z}{dz}=-4 \pi G \rho,\ p= c_s^2 \rho,
\end{equation}
subject to boundary conditions
$g_z$($z=0$)$=0$, $\rho$($z=0$)$=\rho_0$, $p$($z=0$)$=\rho_0c_{s0}^2$, where $\rho_0$ and $c_{s0}$ are the initial density and the sound speed at $z=0$. The initial magnetic field is assumed to be uniform along the \textit{z}-direction: $B_z=B_0 ,\ B_x=B_y=0$, where $B_0 $ is a constant. The simulation of the equilibrium gas sheet is started with random velocity perturbations ($v_x = v_a R_m$($x,y$)$,\ v_y = v_a R_m$($x,y$)$,\ v_z=0$) at each grid point where $R_m$ is a random number with a spectrum $v_k^2 \propto k^n$ in Fourier space and $n$ is either $-4$ or $0$. These correspond to turbulence or white noise, respectively. The turbulence is not replenished, and therefore allowed to decay freely. 
 We use periodic boundaries in the $ x-$ and $y- $ directions and a mirror-symmetric boundary condition at $z = 0$. The computational region is  $-4\pi H_0< x,y < 4\pi H_0$ and $ 0< z <4 H_0$. The number of grid points in each direction is ($N_x, N_y, N_z$) = ($256,256,20$).

\subsection{Numerical Parameters}
As units of length, velocity and density we choose $H_{0} = c_{s0}$/$\sqrt{2\pi G \rho_0}$, $c_{s0}$ and $\rho_{0}$, respectively. This naturally gives the unit of time $t_0 \equiv H_{0}$/$c_{s0}$. The ratio of the initial gas to magnetic pressure at $z=0$ introduces one dimensionless parameter,
\begin{equation}
\beta_{0} = \frac{8 \pi \rho_{0}c_{s0}^{2}}{B_{0}^{2}} .
\end{equation}  
The parameter $\beta_{0}$ is related to the normalized mass-to-flux ratio $\mu_S \equiv 2\pi G^{\frac{1}{2}}\Sigma_S$/$B_0$ for Spitzer's self-gravitating cloud \citep{spi42}, in which $\Sigma_S = 2 \rho_0 H_0$. Therefore,  
\begin{equation}
\beta_{0}=\mu_S^{2}.
\end{equation}  
Dimensional values of all the quantities can be found by specifying appropriate values for $\rho_{0}$ and $c_{s0}$. For example, if $c_{s0} = 0.2~\rm{km ~s^{-1}}$ and $n_{0} \equiv \rho_{0}$/$m_{n}=10^4~\rm{cm}^{-3}$ where $m_{n}=2.33 \times 1.67\times 10^{-24} \rm{g}$, we obtain $H_{0}\simeq 0.05~ \rm{pc}$, $t_{0}\simeq 2.5\times 10^{5}~ \rm{yr}$ and $B_{0} \simeq 40 ~\mu \rm{G}$ if $\beta_{0} = 0.25$. The unit of column density is $\Sigma_0=\rho_{0} H_{0} \simeq 6 \times 10^{-3} \rm{g}\, cm^{-2}$, which corresponds to a number column density $N_0 \equiv \tilde{\Sigma}_0$/$m_n \simeq 1.5 \times 10^{21}  \rm{cm^{-2}}$. We define $\tilde{\Sigma} = \Sigma$/$ \Sigma_{0}$ as the normalized column density.

\begin{table}
	\centering
	\caption{$\beta_{0} $ is the initial ratio of thermal to magnetic pressure at $z = 0$, $v_a$ is the amplitude of the initial velocity fluctuation, $t_{\rm core}$ is the time needed for core formation.}
	\label{Parameter}
	\begin{tabular}{lccrrrr} 
		\hline
		\hline
		Model & $\beta_0$ & Spectrum & ${v_a}$/${c_s}$ & $t_{\rm{core}}$/$t_0$ & Comments  \\
		\hline
		V1 &  0.25 & $k^{-4}$ & 0.1 &   87.6      \\
		V4 &  0.25 & $k^{-4}$ & 3.0  & 16.9&Fiducial model\\
		K1 &  0.25 & $k^{0}$  & 3.0  & 95.7  \\
		B3 &  4.0  & $k^{-4}$ & 3.0  & 1.13 & Initially supercritical\\
		B4 &  9.0  & $k^{-4}$ & 3.0  & 1.06& Initially supercritical\\
		B8 &  4.0  & $k^{-4}$ & 0.1  & 7.36 & Initially supercritical  \\
		\hline
	\end{tabular}
\end{table}
\section{Results}\label{results}

Table \ref{Parameter} summaries the simulation results for all the models and for different parameters. In the table we have listed the values of $\beta_{0}$, the form of power spectrum, $v_{k}$, and the amplitude of the initial velocity fluctuation $v_{a}$. We have also listed the core formation time $t_{0}$, which is defined as the time when the maximum density of a core reaches $100\, \rho_{0}$. Depending on the value of the initial mass-to-flux ratio, we have classified the models as subcritical ($\beta_{0}< 1$) or supercritical ($\beta_{0}> 1$). Model V1 and V4 are subcritical clouds, in which we have changed the amplitude of the initial velocity fluctuation $v_{a}$ but with the turbulent spectrum fixed at $v_k^2 \propto k^{-4}$. Model K1 is also subcritical and has an initial velocity fluctuation $v_a = 3.0\,c_{s0}$ but the power spectrum is white noise, $v_k^{2} \propto k^{0}$. In the models B3 to B8, we have $\beta_{0} > 1$, and the cloud is supercritical with initial velocity fluctuation in both supersonic ($v_a = 3.0\, c_{s0}$) and subsonic ($v_a = 0.1\, c_{s0}$) limits. We are particularly interested in the column density PDF of the subcritical clouds where the magnetic support prevents rapid gravitational collapse. Instead, the cloud oscillates and settles into a quasiequilibrium state of filamentary structure due to the interplay of turbulence, magnetic support and gravitationally driven ambipolar diffusion (see \cite{aud16} for details). Interestingly, the subcritical clouds with linear perturbations tend to have a much steeper slope in their column density PDF ($\alpha \simeq 4 $) as discussed later in Section~\ref{subcritical}.

\begin{figure}
	\includegraphics[width=\columnwidth,trim=52mm 90mm 60mm 90mm, clip=true]{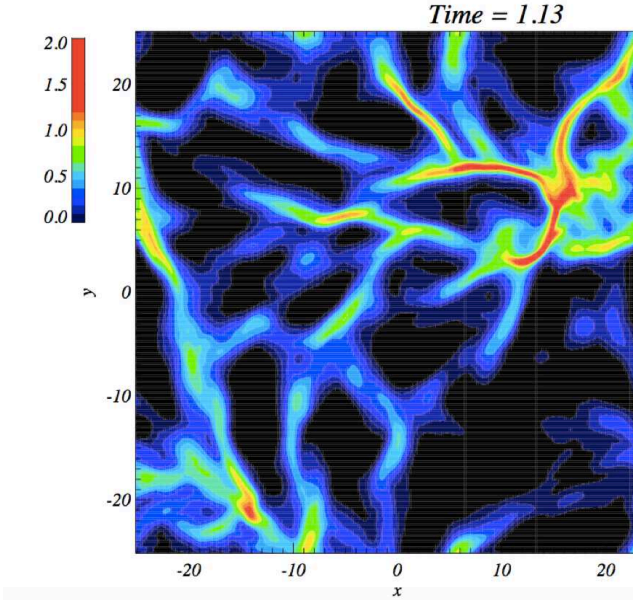}
    \caption{Logarithmic column density ($\tilde{\Sigma} = \Sigma$/$\Sigma_0$) contours at $t=1.13 \, t_{0}$ for the model B3 with nonlinear velocity spectrum $v_{k}^2 \propto k^{-4}$ of amplitude $v_{a}= 3.0\, c_{s0}$. The model B3 is initially supercritical (i.e. $\beta_{0} = 4$). The $x$ and $y$ axes are in the units of $H_{0} \simeq 0.05$ pc. The maximum column density is located at ($x,y$)$ = $($15.4 \, H_{0}, 8.3 \, H_{0}$).  The unit of time is $t_{0} \simeq 2.5 \times 10^{5}$ yr. The panel shows the column density when viewed face on along the direction of the magnetic field ($z-$axis).}
    \label{B3_clm}
\end{figure}

\subsection{General properties of the supercritical cloud}
Here we discuss models that are initially supercritical with $\beta_0 > 1$. We consider a velocity spectrum $v_{k}^{2} \propto k^{-4}$ with both linear and nonlinear initial velocity amplitude. Figure~\ref{B3_clm} shows the time snapshot of the logarithmic column density at the end of the simulation for model B3 ($t= 1.13 \, t_{0} $). This model starts with an initially supercritical mass-to-flux ratio with $\beta_{0} = 4$.  The figure shows the column density in the $x-y$ plane at the end of the simulation when the maximum density is $100 \, \rho_{0}$. The column density is obtained by integrating the sheet along the direction of the magnetic field ($z$ axis). We assume that the cloud is viewed face on along the short axis whose width is typically set by the hydrostatic equilibrium  along the magnetic field.

\begin{figure}
	\includegraphics[width=\columnwidth, trim=52mm 90mm 60mm 90mm, clip=true]{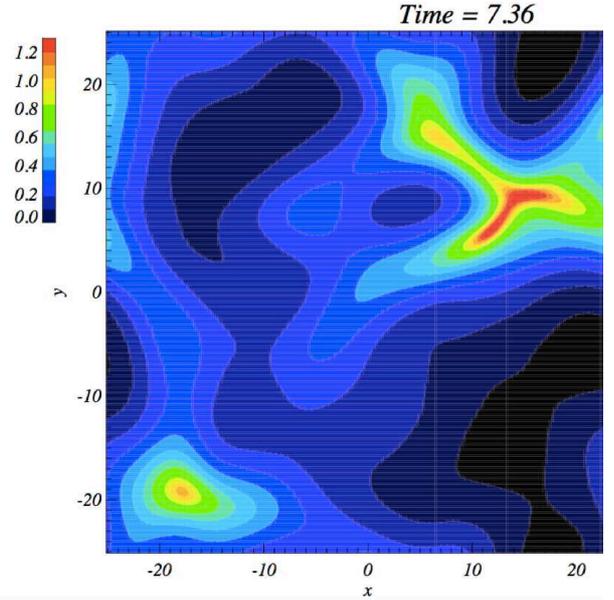}
	\caption{Logarithmic column density contours at $t=7.36 \, t_{0}$ for the model B8 with linear initial velocity amplitude $v_{a}= 0.1\, c_{s0}$. All the other parameters are the same as model B3.  The panel shows the column density when viewed face on along the direction of the magnetic field ($z-$axis).}
    \label{B8_clm}
\end{figure}

Figure~\ref{B8_clm} shows the time snapshot of the logarithmic column density for model B8 at $t= 7.36 \, t_{0} $. Model B8 corresponds to a linear perturbation with initial velocity amplitude $v_{a} = 0.1 \, c_{s0}$. All the other parameters are the same as model B3. The core formation time for model B3 ($t=1.13 \, t_{0}$) with nonlinear velocity perturbation is much less than for model B8 ($t=7.36 \, t_{0}$) with a linear perturbation. Furthermore, visual inspection of Figures~\ref{B3_clm} and \ref{B8_clm} shows that the column density distribution for model B3 is much more filamentary than for model B8. The filaments in model B3 are distributed throughout the simulation region, with some of them having more condensed regions with higher column density. The maximum column density is located at ($x,y$)$ = $($15.4 \, H_{0}, 8.3 \, H_{0}$). Model B8 evolves much more slowly until one of two dense regions go into a runaway collapse due to gravity. The maximum column density for model B8 is located at ($x,y$)$ = $($14.8 \,H_{0}, 9.3 \, H_{0}$). The denser filamentary network in model B3 can be attributed to the initial velocity amplitude that is 30 times greater than that in model B8.  As the clouds are supercritical to begin with, the large-scale supersonic turbulence condenses the gas and gravity causes the dense regions to go into a runaway collapse. In model B4 we further decrease the strength of the magnetic field, keeping all the other parameters similar to model B3. The intrinsic nature of the column density remains the same only with a tiny difference in the core formation time as shown in Table~\ref{Parameter}. The presence of a weak magnetic field does not seem to have a strong impact compared to the decaying turbulence and gravity in the evolution of these supercritical clouds. 
\begin{figure}
	\includegraphics[width=\columnwidth]{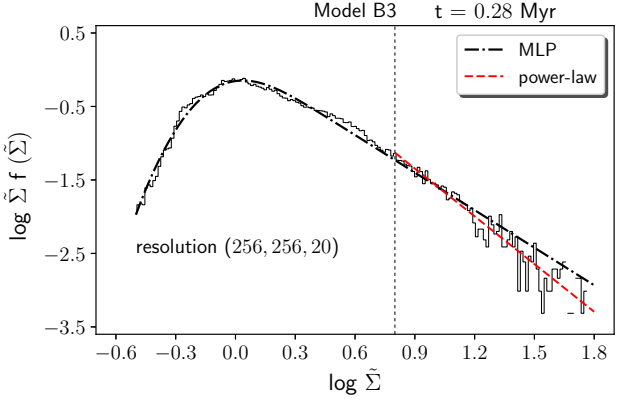}
    \caption{The column density PDF of model B3 at $t = 0.28~ \rm{Myr}$.  The dashed-dotted line is the best-fit MLP distribution with parameters $\alpha=1.7$, $\mu_{0}=-0.25$, and $\sigma_{0}=0.37$. The vertical axis is $\log \tilde{\Sigma} f$($\tilde{\Sigma}$)$= \log $($\Delta N^{\prime}$/$\Delta \log \tilde{\Sigma}$), where $\Delta N^{\prime}=\Delta N $/($N_{\rm{total}} \ln\left(10\right) $)  and $ \Delta \log \tilde{\Sigma} = 0.01$. The dashed red line in the best-fit power law to the tail of the column density PDF ($\log \tilde{\Sigma} \ge 0.8$) with power-law index $\alpha = 2.2$. All fits are done using maximum likelihood estimation and are independent of binning. }
    \label{B3}
\end{figure}

\subsubsection{The Column Density PDFs for supercritical clouds}
\begin{figure}
	\includegraphics[width=\columnwidth]{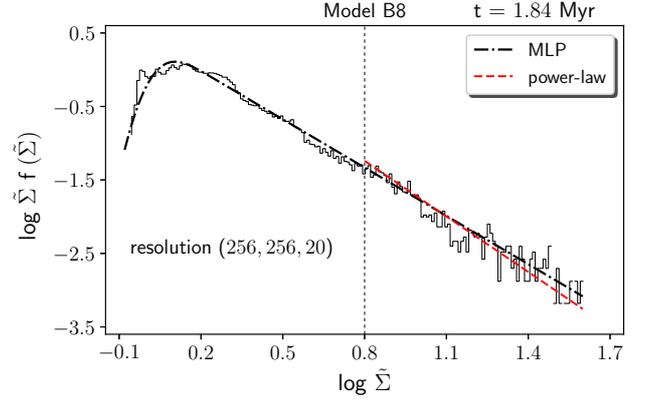}
    \caption{ The column density PDF of model B8 at $t =1.84~ \rm{Myr}$. The dashed-dotted line is the best-fit MLP distribution with parameters $\alpha=2.2$, $\mu_{0}=0.06$, and $\sigma_{0}=0.14$. The dashed red line in the best-fit power law to the tail of the column density PDF ($\log \tilde{\Sigma} \ge 0.8$) with power-law index $\alpha = 2.5$. }. 
    \label{B8}
\end{figure}

Figures~\ref{B3} and \ref{B8} show the column density PDFs along with the best-fit modified lognormal power law (MLP) and Pareto distributions \citep{cla09}, for the two models B3 and B8, respectively. The MLP distribution is a three-parameter PDF given in closed form as
\begin{equation}
\begin{split}
f\left({\Sigma}\right)&= \frac{\alpha}{2}\exp\left(\alpha \mu_{0}+\frac{\alpha^{2}\sigma_{0}^{2}}{2}\right)\,\Sigma^{-\left(1+\alpha\right)}  \\
& \times~  \rm{erfc} \left(\frac{1}{\sqrt{2}}\left(\alpha \sigma_{0} - \frac{\ln \Sigma - \mu_{0}}{\sigma_{0}}\right)\right)~, ~\Sigma \in \left[0,\infty\right)
\end{split}
\end{equation}
\citep{bas15}. Here $\Sigma$ is the column density of the molecular cloud, and the three parameters describing the MLP distribution are $\alpha$, $\mu_{0}$ and ${\sigma}_{0}$. The power-law tail is represented by $\alpha$, while $\mu_{0}$ and $\sigma_{0}$ describe the body of the distribution (see \cite{bas15} for details).  
Here, we find the set of parameters for the MLP distribution that fits the normalized column density PDF. The relative similarity in the normalized column density PDFs for both the models is evident in the fit parameters. Figure \ref{B3} shows the normalized column density PDF of model B3 obtained at the end of the simulation at $t =1.13\,t_{0}$, i.e. $0.28$ Myr. Here we plot $\log $($\Delta N^{\prime} $/$\Delta \log \tilde{\Sigma}$)$ \equiv \log \tilde{\Sigma} f\left(\tilde{\Sigma}\right) $, where $\Delta N^{\prime} = \Delta N $/($N_{\rm{total}} \ln\left(10\right)$)  and $ \Delta \log \tilde{\Sigma} = 0.01$ with $\log \tilde{\Sigma}$ in the horizontal axis. We fit the MLP distribution using the maximum likelihood estimation \citep{jon02}.  The best-fit parameters  are $\alpha=1.7 $, $\mu_{0}=-0.25 $, $\sigma_{0} = 0.37$.  For all the fitting routines  we have used the \textit{PYTHON} optimization module \textit{scipy.optimize.differential.evolution} and \textit{scipy.optimize.basinhopping} to find the global minimum of the function. Figure \ref{B8} shows the normalized column density PDF of model B8 obtained at the end of the simulation at $t = 7.36\,t_{0}$, i.e. $1.84$ Myr. Similarly, we fit the MLP  to $\tilde{\Sigma} f$($\tilde{\Sigma}$) using the maximum likelihood estimation (MLE) method. We note that the MLE fits are independent of binning. The best-fit parameters are $\alpha=2.2 $, $\mu_{0}=0.06 $, and $\sigma_{0} = 0.14$. 

For both the models B3 and B8 we also fit the Pareto distribution ($ f\left(\Sigma\right)\propto \Sigma^{-\left(1+\alpha\right)}$) to the tail of the column density PDF, i.e. where $\log \tilde{\Sigma} \ge 0.8$. In Figures \ref{B3} and \ref{B8}, the dashed red line represents the best-fit power law using MLE, yielding $\alpha = 2.2 $ and $\alpha =2.5$  for model B3 and B8, respectively. The power-law fits are slightly steeper than the MLP fits. This is because the MLP gives a global fit to the entire distribution, including the turnover at the lower column density values. Thus, depending on the turnover point the MLP fits adjust accordingly and give a global representation of the entire distribution. 
We performed several simulations with same initial conditions but different (increased) spatial resolution, different random perturbation seeds, and different fitting routines, and found that the value of the measured power-law index $\alpha$ has a variability of $10\% - 20\%$.  The slightly steeper slope for model B8 can be attributed to the smaller initial velocity perturbation. There are fewer regions with dense gas compared to model B3 as evident in Figures~\ref{B3_clm} and \ref{B8_clm}. The column density PDF for model B3 with nonlinear perturbations has a wider spread (indicated by $\sigma_{0}$) compared to model B8 with linear perturbations. More importantly, for both these supercritical models, the power law establishes itself at a very early stage of evolution so there is no significant observable time with a lognormal PDF. The power law dominates almost the entire distribution since many regions go into direct collapse, i.e. there is a global fragmentation and gravitational collapse. 

\begin{figure}
	\includegraphics[width=\columnwidth, trim=52mm 90mm 60mm 90mm, clip=true]{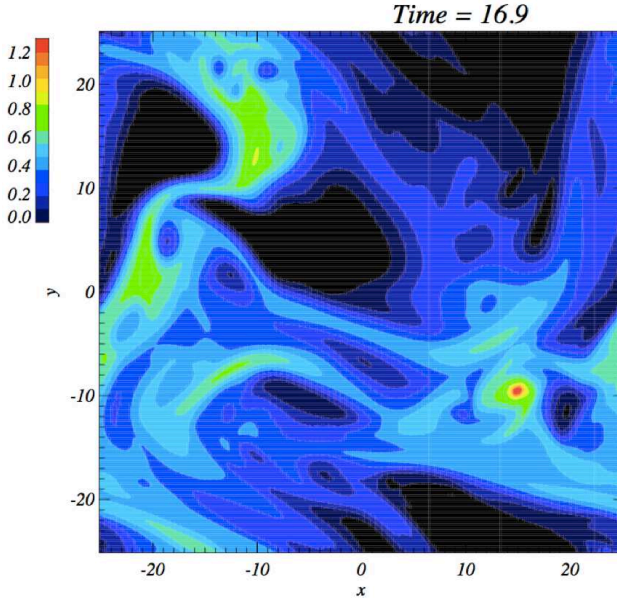}
    \caption{Logarithmic column density contours at $t=16.9 \, t_{0}$ for the model V4. The model V4 is the fiducial model, in which the turbulent spectrum is $v_{k}^{2} \propto k^{-4}$ with the initial velocity amplitude $v_{a}= 3 \, c_{s0}$. This is a subcritical cloud with the initial mass-to-flux ratio of about 0.5 (i.e. $\beta_{0} = 0.25$). The unit of time is $t_{0} \simeq 2.5 \times 10^{5} $ yr. The figure shows the column density when viewed face on along the direction of the magnetic field ($z-$axis). }
    \label{V4_clm}
\end{figure}

\subsection{General properties of the subcritical cloud}   \label{subcritical}
We discuss the result of model V4 as a fiducial model, where the cloud has $\beta_{0} = 0.25$, corresponding to a normalized mass-to-flux ratio of about $0.5$. We initiate the simulation with an initial nonlinear turbulent velocity perturbation of amplitude $v_{a} = 3.0\, c_{s0}$.  
Figure \ref{V4_clm} shows the time snapshot of the logarithmic column density colour map of model V4. The snapshot is obtained at the end of the simulation, when the maximum density is $100\, \rho_0$. The figure shows the column density in the $x-y$ plane that is obtained by integrating the sheet along the direction of the magnetic field ($z-$axis).

\begin{figure}
	\includegraphics[width=\columnwidth, trim=52mm 90mm 60mm 90mm, clip=true]{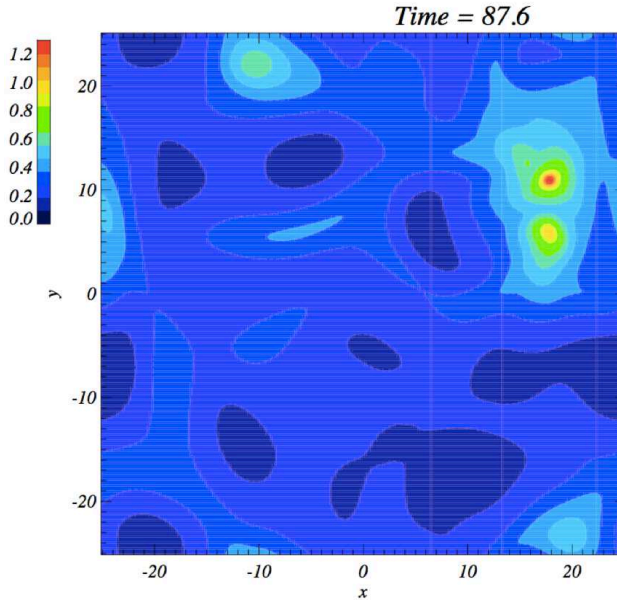}
    \caption{Logarithmic column density contours at $t=87.6 \, t_{0}$ for the model V1 corresponding to the initial velocity amplitude $v_{a}= 0.1 \, c_{s0}$. This is also a subcritical cloud (similar to ModelV4) with the initial mass-to-flux ratio of about 0.5 (i.e. $\beta_{0} = 0.25$) with the turbulent spectrum is $v_{k}^{2} \propto k^{-4}$. The unit of time is $t_{0} \simeq 2.5 \times 10^{5} $ yr. The figure shows the column density when viewed face on along the direction of the magnetic field ($z-$axis).}
    \label{V1_clm}
\end{figure}

\begin{figure}
	\includegraphics[width=\columnwidth,trim=52mm 90mm 60mm 90mm, clip=true]{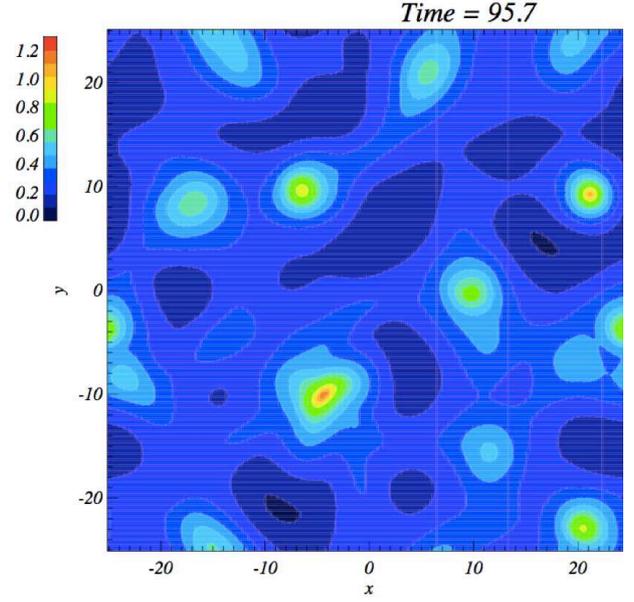}
    \caption{Logarithmic column density contours at $t=95.7 \, t_{0}$ for the model K1. The model K1 has a turbulent spectrum $v_{k}^{2} \propto k^{0}$. All other initial conditions are same as the fiducial model V4. The initial velocity amplitude $v_{a}= 3 \, c_{s0}$ and initial mass-to-flux ratio is about 0.5 (i.e. $\beta_{0} = 0.25$). The unit of time is $t_{0} \simeq 2.5 \times 10^{5} $ yr. The figure shows the column density when viewed face on along the direction of the magnetic field ($z-$axis). }
    \label{K1_clm}
\end{figure}

\begin{figure*}
\begin{center}
	\includegraphics[width=7.5in,trim=13mm 0mm 7mm 8mm, clip=true]{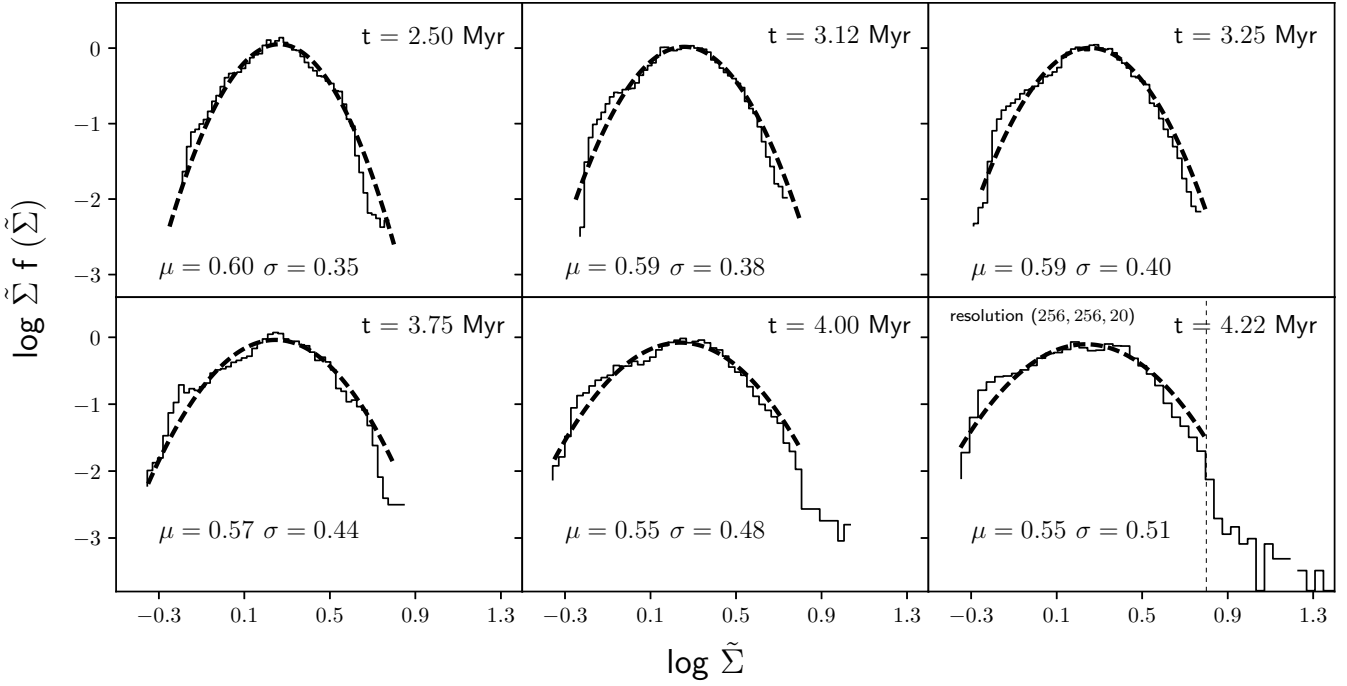}
    \caption{The time evolution of the column density PDF of the fiducial model V4 along with the best-fit lognormal distribution in the normalized form. The time corresponding to each snapshot is indicated on the top of each panel. The value of the fit parameters $\mu$ and $\sigma$ are also shown on the bottom left corner on individual panels. The dotted black line parallel to the $y-$axis at $\log_{10} \tilde{\Sigma} = 0.8$ on the final subplot marks the power-law zone.}
    \label{V4_evol}
\end{center}
	\end{figure*}

Figure~\ref{V1_clm} shows the time snapshot of the logarithmic column density at the end of the simulation for model V1. The model V1 corresponds to linear perturbations with $v_{a}=0.1 \, c_{s0}$. All the other parameters are same as the fiducial model V4. One of the main differences between model V4 and V1 is the core formation time. It is $t = 16.9 \, t_{0}$ for model V4 but a much longer time $t= 87.6\, t_{0}$ for model V1 with its initial linear perturbation. By visual inspection of Figures~\ref{V4_clm} and \ref{V1_clm}, one can also see the morphological difference in the distribution of dense structures. The fiducial model is much more filamentary and has more evolved collapsing cores within the vicinity of the maximum column density located at ($x,y$) = ($14.8\, H_{0}, -9.5\, H_{0}$). Although model V1 has evolved much longer than model V4, the weak velocity perturbation causes very little compression of the gas. Furthermore, the presence of strong magnetic support prevents the gas from collapsing due to self-gravity. The dense regions settle into an oscillatory equilibrium between magnetic and gravitational forces, with the neutrals gradually diffusing through the field lines due to ambipolar diffusion and forming denser regions. 

Figure~\ref{K1_clm} shows a time snapshot of the column density of model K1 at the end of the simulation. Model K1 has a turbulent spectrum $v_{k}^{2} \propto k^{0}$ with all the other parameters same as the fiducial model. Although a flat spectrum is not consistent with observations, we use it to compare with the fiducial model. The white noise turbulence creates compressions in the initial stages that are more localized than in model V4 and they are not filamentary. Eventually, ambipolar diffusion leads to the formation of widely distributed cores. The core formation time ($t = 95.7 \, t_{0}$) is much longer than in the fiducial model. Furthermore, the cores are mostly circular in shape with a lack of filamentary structures.

\subsubsection{The column density PDFs for subcritical clouds} \label{subcritcalPDF}
Figure~\ref{V4_evol} shows the time evolution of the column density PDF of the fiducial model V4. Here we plot $\log $($\Delta N^{\prime} $/$\Delta \log \tilde{\Sigma}$)$ \equiv \log \tilde{\Sigma} f$($\tilde{\Sigma} $) and show the snapshots at different times of the column density starting from $t=10\,t_{0} $, (i.e. $t = 2.50$ Myr) till the end of the simulation when the maximum density is $100\, \rho_{0}$. The time corresponding to each snapshot is indicated on the top of each panel.  In each panel, the PDF is overplotted with the best-fit lognormal distribution 
\begin{equation}
	f\left(\Sigma\right) = \frac{1}{\Sigma \ \sqrt{2\pi}\sigma}\exp\left(-\frac{\left(\ln \Sigma - \mu\right)^2}{2\sigma^2}\right) 
	\label{lognormal}
\end{equation}
(till the cutoff point $\tilde{\Sigma} = 6.3$). We fit the lognormal distribution using MLE and the fit parameters $\mu$ and $\sigma$ of the successive epochs during the time evolution are shown in each panel.

The first panel on the top left of Figure~\ref{V4_evol} is a snapshot of the column density PDF at a very early stage of evolution. The PDF is predominantly lognormal with a broad spread about its mean. A best-fit lognormal distribution (black dashed line) has the parameter values $\mu = 0.60$ and $\sigma =0.35$. The lognormal shape in the early stages can be attributed to the initial nonlinear perturbation \citep{sem94} (discussion in the next section). As the cloud evolves further, it gets compressed due to the large-scale flow and develops some pockets of high column density. Then it rebounds and shows oscillation. Thus with each successive compression more regions with high column density develop. The maximum density is also strongly increased during each compression (see figure 14 in \cite{kud11}) due to the supersonic flow. This feature is very pronounced in Figure~\ref{V4_evol} where we see the width of the column density PDF gradually widens over time (see also \cite{war14,tas10}), hence it is not just a fixed value solely depending on the initial strength of the nonlinear perturbation (supersonic turbulence)\footnote{Although the width is growing as turbulence decays (decreasing Mach number), the final column density PDF (see section \ref{3.2.2}) is consistent with the trend found in hydrodynamic simulations with driven turbulence (e.g., \cite{fed13}), where greater initial Mach number corresponds to a wider PDF.}. This is shown in the increase in the value of the lognormal fit parameter $\sigma$.  
After several oscillations, the local pockets of higher column density become supercritical and go into a runaway collapse. The column density PDF retains its lognormal shape for the most part of the evolution. However, at around $t = 4.00~\rm{Myr}$, it builds up regions of high column density. A distinct power-law tail gradually emerges, as seen in the bottom row of Figure~\ref{V4_evol}.

\begin{figure}
	\includegraphics[width=\columnwidth]{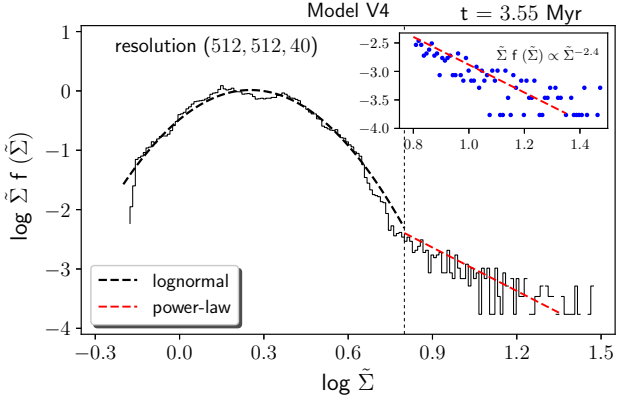}
	\caption{The column density PDF of model V4 at $t = 3.55~ \rm{Myr}$ with spatial resolution (512,512,40). The dashed black line is best-fit lognormal distribution using the maximum likelihood estimation and has parameters $\mu_0=0.58$ and $\sigma_0=0.39$. The dashed red line represents the best-fit power law using maximum likelihood with power-law index $\alpha = 2.4$. The two regions are separated by a dotted black line at $\log\tilde{\Sigma} = 0.8$. The inset box on the top right corner is a zoomed in view of the power-law region along with the best-fit line.}
    \label{V4_512}
\end{figure}

To resolve the high density regions and highlight the power law at the final snapshot,
%
%
we further perform a simulation with same initial conditions as model V4 but with twice the spatial resolution. Therefore ($N_{x},N_{y}, N_{z}$) = ($512,512,40$), and we follow the simulation till the maximum density is $100\, \rho_{0}$. As reported previously in \cite{kud11}, the core formation time becomes slightly shorter for the high-resolution cases. In the case of model V4 the core formation time is $t = 14.2\, t_{0}$. It should be noted that the realization of the random perturbation to the initial velocity fluctuations are also not the same when adopting a different resolution (see \cite{kud11} for details). Figure~\ref{V4_512} shows the best-fit power law (red dashed line) along with the best-fit lognormal distribution (black dashed line) for the fiducial model V4 but for a higher spatial resolution ($512,512,40$). The zoomed-in inset box on the upper right corner in Figure~\ref{V4_512} shows the power-law fit in the $\log \tilde{\Sigma} f$($\tilde{\Sigma}$) vs $\log \tilde{\Sigma}$ plot, which has an index $\alpha = 2.4$.
The gradual development of a power-law tail in the later stages of the evolution can be attributed to gravitationally-driven ambipolar diffusion occurring within the compressed filaments (see \citet{kud14} for an analytic model). The neutrals drift past the field lines to form supercritical pockets within the filaments, on a somewhat shortened ambipolar diffusion time scale, and rapid collapse ensues in those regions. 

\begin{figure}
	\includegraphics[width=\columnwidth]{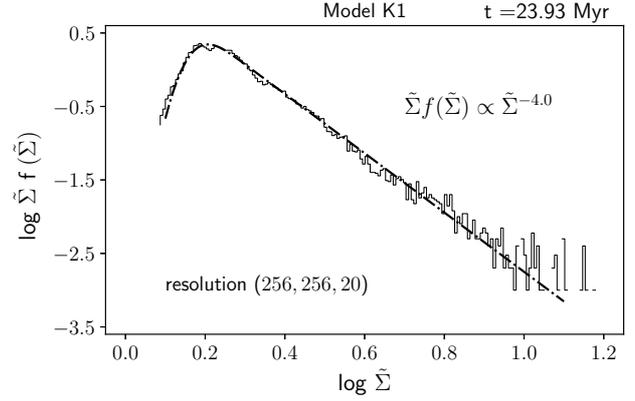}
    \caption{MLP fit to the column density PDF of model K1 at $t = 23.93~ \rm{Myr}$. The best-fit parameter values are $\alpha=4.0$, $\mu_{0}=0.37$, and $\sigma_{0}=0.09$.}
    \label{K1}
\end{figure}

Figure~\ref{K1} shows the normalized column density PDF of model K1 along with the best-fit MLP distribution. The best-fit parameters are $\alpha=4.0 $, $\mu_{0}=0.37 $, $\sigma_{0}= 0.09$. This model has a flat spectrum $v_k^2 \propto k^0$, but is otherwise the same as the fiducial model V4. The white noise turbulence creates compressions in the initial stages that are more localized than in model V4 and they are not filamentary. Eventually, ambipolar diffusion leads to the formation of widely distributed cores. The formation process of the cores is ultimately more similar to that of model V1 (linear perturbations) discussed below, as there are no large scale density compressions that create a filamentary structure (see also \citet{bas09}). An ensemble of cores formed through ambipolar diffusion is expected to have a very steep column density PDF as explained in Section~\ref{discussions}.



\subsubsection{Effect of Initial Velocity Amplitude}\label{3.2.2}
\begin{figure}
	\includegraphics[width=\columnwidth]{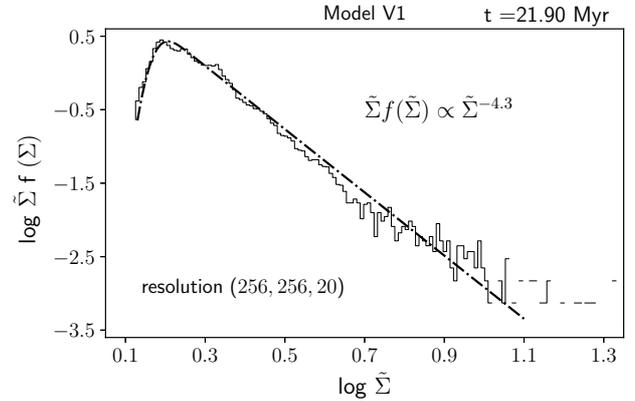}
    \caption{The column density PDF of model V1 at $t =21.90~ \rm{Myr}$. The dashed-dotted line is the best-fit MLP distribution with parameters $\alpha=4.3$, $\mu_{0}=0.40$, and $\sigma_{0}=0.06$. }
    \label{V1}
\end{figure}

Model V1 has same initial conditions as our fiducial model V4 but with varying velocity perturbation amplitude $v_{a}$. The characteristics of the column density PDF of this model are studied using the best-fit MLP distribution. 
Figure \ref{V1} shows the normalized column density PDF of the model V1 obtained at the end of the simulation at $t = 87.6\,t_{0}$, i.e. $21.90$ Myr. We fit the MLP distribution to $\tilde{\Sigma} f$($\tilde{\Sigma}$) using MLE. The best-fit parameters are $\alpha=4.3 $, $\mu_{0}=0.40$, $ \sigma_{0} = 0.06$. 



With the best-fit MLP distribution, it is relatively easy to quantify several important properties of the column density PDF for the subcritical cloud with different initial velocity fluctuations. 
For both the models, the magnetic field provides the dominant support against collapse in the initial subcritical phase. The fiducial model V4 with nonlinear turbulence goes through an oscillatory filamentary phase, with magnetic pressure and tension acting like a spring working against the initial compression. 

A major part of the evolution is dominated by a lognormal distribution as seen in Figure~\ref{V4_evol}. However, model V1 does not show any oscillations. It settles into a quasiequilibrium between gravitational and magnetic forces and evolves rather slowly compared to model V4. While Figure \ref{V4_512} shows that the power law is prominent only at higher column densities ($\log \tilde{\Sigma} \geq 0.8$) in model V4, for model V1 the power law is not only steep but extends all the way to much lower values of column density as shown in Figure~\ref{V1}.  The value of $\alpha$ is relatively high in the $\log $($\tilde{\Sigma} f$($\tilde{\Sigma}$))$\,\rm{vs}\,\log{\tilde{\Sigma}}$ plot. Another obvious distinction between the two PDFs is the spread of the distribution indicated by the values of the fit parameters $\sigma$ and $\sigma_{0}$. The fiducial model has a much broader distribution ($\sigma = 0.39$) compared to the model V1 ($\sigma_{0} = 0.06$). The difference in the spread of the PDF between the models is a natural imprint of the differing initial velocity perturbation amplitude $v_{a}$. The supersonic turbulent initial condition for model V4 results in a wider spread of the distribution compared to model V1.


\subsection{Comparison of PDFs}

\begin{figure}
	\includegraphics[width=\columnwidth]{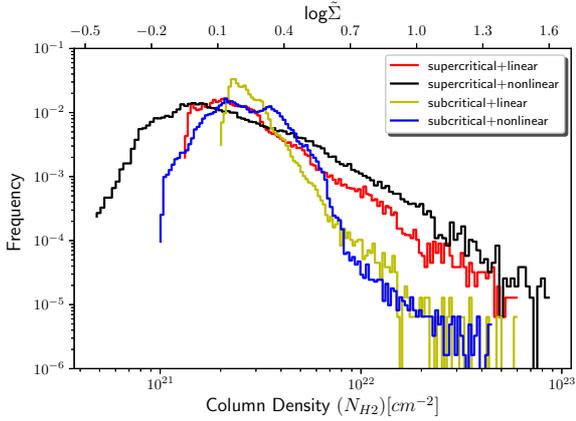}
    \caption{The column density PDFs of simulated models of molecular cloud with different initial conditions. The red and the black histogram are  the supercritical model B8 and B3 with linear and nonlinear perturbation respectively. The yellow and the blue histogram are both subcritical clouds with linear (model V1) and nonlinear perturbation (model V4) respectively. The vertical axis is the normalized frequency  $\Delta N^{\prime} = \Delta N $/($N_{\rm{total}} \ln$($10$)) with the data binned with a uniform spacing of $\Delta \log \tilde{\Sigma} = 0.018$. Blue histogram is the high resolution (512,512,40) version of model V4.}
    \label{comparison} 
\end{figure}

Figure~\ref{comparison} is a composite plot that shows the column density PDFs of simulated molecular clouds with four different initial conditions. The vertical axis is the normalized frequency $\Delta N^{\prime} = \Delta N $/($N_{\rm{total}} \ln$($10$)) with the data binned with a uniform spacing of $\Delta \log \tilde{\Sigma}$.  The horizontal axes are $\log \tilde{\Sigma}$ on the top and number column density $N_{\rm{H2}}$ in $\rm{cm}^{-2}$ at the bottom, where $N_{\rm{H2}} = \tilde{\Sigma} N_{0}$. The column density PDF for clouds with supercritical mass-to-flux ratio (red and black histogram) is mostly power law above $\sim 10^{21} \,\rm{cm}^{-2}$. In contrast, models with strong magnetic field show different characteristics. The column density PDF of the subcritical model V4 (blue histogram) is predominantly lognormal with a power-law tail above $ \approx 9 \times 10^{21}\, \rm{cm}^{-2}$. Furthermore, for model V1 the column density PDF is primarily a power law (no lognormal body) like the supercritical cases but with a much steeper slope. We summarize the fit parameters of the MLP function for all models (except V4) in Table~\ref{MLPfit parameter}.  \\

\begin{table}
	\centering
	\caption{Fit parameters for the MLP distribution to the column density PDF for models with different initial conditions.}
	\label{MLPfit parameter}
	\begin{tabular}{lccrrrr} 
		\hline
		Model  & $\alpha$ &$\mu_{0}$ & $\sigma_{0}$  \\
		\hline
		V1 &  4.3 & 0.40	&0.06 \\
		K1 &  4.0    & 0.37      & 0.09 \\
		B3 &  1.7 &-0.25	&0.37\\
		B4 &	  1.8 & -0.55  & 0.17\\
		B8 &  2.1 & 0.06	&0.14\\

		\hline
	\end{tabular}
\end{table}

\section{Discussion}\label{discussions}

What role does a strong magnetic field play in shaping the column density PDFs and controlling star formation? Many theorists \citep{kri11,fed13} assert that an increasing magnetic field strength acts as an extra cushion against turbulent compression but has no definite influence on the power-law slope of the PDFs. Some indicate a weak steepening of the power-law tail with increasing magnetic field strength \citep{col12}.  
	Our models explore a larger dynamic range of magnetic field strength and instead
demonstrate a clear trend where subcritical models (strong magnetic field) with linear perturbations show minimum star formation and have a steep power-law tail with index $\alpha \approx 4$. This broadly agrees with recent observational results from Herschel and Planck \citep{lom15,alv17}, which show that quiescent clouds with reduced star formation have similar power-law features. Our results contradict the paradigm that clouds with little star formation have a lognormal PDF. 

Furthermore, it is often debated whether a lognormal PDF is a direct imprint of supersonic turbulence alone, with column density PDFs becoming wider with increasing Mach number  \citep{col12,fed13}  and narrower with increasing magnetic field strength \citep{mol12}. However, only our fiducial model V4 (with both strong magnetic field and supersonic turbulence) mimics the results of these simulations (of magnetized clouds with driven turbulence) where the PDF is predominantly lognormal with a power law at the high density end. 
The column density map of this model (Figure \ref{V4_clm}) is highly filamentary and oscillating due to the interplay between the magnetic and ram pressures of the large-scale flow, as described analytically by Auddy et al. (2016).
The presence of a strong magnetic field acts like a physical spring against the turbulent compression, and causes a sequence of oscillations. This minimizes the decay of turbulence as most of the energy stays on the large scale for an extended time (resembling driven turbulence) resulting in a nearly lognormal PDF until gravitationally-driven ambipolar diffusion takes over. 
In all other models, the column density PDF develops a power law with a peak and turnover at lower values. These PDFs should not be classified as lognormal just because they have a peak. Indeed observations often show a pure power-law PDF for both star forming and diffuse clouds \citep{lom14,lom15,alv17}. One possible explanation \citep{war14} is that as the cloud evolves, the underlying lognormal shape may be lost. Our results strongly indicate that magnetic support in unison with turbulence and gravity play a crucial role in shaping the different observed PDFs.


A physical interpretation of these results requires understanding the fragmentation process in magnetically supported clouds. 
For example, in the model of \cite{bas94}, the gravitational contraction of supercritical cores embedded in a subcritical envelope occurs in a very nonhomologous manner. The ultimately self-similar evolution results in supercritical cores having a central near uniform-column density surrounded by a radial power-law profile proportional to $r^{ -1}$ (\citet{bas97}; and see Equation \ref{disk}). This is the same as for hydrodynamic self-similar collapse \citep{lar69,shu77,hun77}. However the convergence to this solution occurs only at innermost radii, while an intermediate region makes the transition from a magnetically supported envelope. Thus, the majority of the core area has a shallower column density profile, and at the boundary the profile of $\Sigma$ is proportional to $r^{-0.7}$ (see Figure 1 in \cite{bas97}). 
As shown in Appendix (\ref{disk geometry}), the radial column density profile with a power-law $\Sigma \propto r^{-1}$ as in Equation (\ref{n2diskpdf}) corresponds to a PDF $\Sigma f$($\Sigma$)$ \propto \Sigma ^{-2}$ but a profile $\Sigma \propto r^{-0.5}$ for example will correspond to a PDF $\Sigma f$($\Sigma$)$ \propto \Sigma ^{-4}$ (Equation (\ref{n1diskpdf})).
The delay in reaching the self-similar collapse in the magnetically supported cloud, as opposed to a non-magnetically-supported cloud is the critical factor that makes the PDF steeper for a magnetically supported cloud.
This is consistent with our high-resolution fiducial model V4, which develops a power-law tail with index $\alpha = 2.4$ in the high column density end ($\log \tilde{\Sigma} \geq 0.8 $)  as shown in Figure~\ref{V4_512}.  The column density PDFs of our simulated models B3 and B8 with supercritical mass-to-flux ratio have a power-law tail with MLP fit parameter $\alpha = 1.7$ and $\alpha = 2.2$ respectively. This is also a direct imprint of the radial power-law profile ($\Sigma\propto r^{-1}$) within the core. In contrast, the column density PDFs of the models V1 and K1 is much steeper with indices $\alpha = 4.3$ and $\alpha = 4.0$, respectively. In both these models with initially small amplitude perturbations, the turbulent energy is confined to small scales, causing primarily local collapse due to gravitational contraction driven by ambipolar diffusion (see Figures~\ref{V1_clm} and \ref{K1_clm}). 
There is a slow transition towards a gravitationally collapsing supercritical inner core from an ambient magnetically dominated regions. Most of the gas is in a transition zone in which the radial column density profile is significantly shallower than $\Sigma \propto r^{-1}$, in fact closer to $\Sigma \propto r^{-0.5}$.
These subcritical models can be directly identified with observations of Polaris and Pipe \citep{lom15}, which exhibit steeper power law with indices $\alpha= 3.9$ and $\alpha=3.0$, respectively. 

\section{Conclusion}\label{conclusion}

We have presented a unified model of column density PDFs that accounts for lognormal plus power law PDFs in one limit and peaked power laws with different indices in other limits. We employed fully three-dimensional magnetohydrodynamical simulations with either supercritical or subcritical mass-to-flux ratio, and including ambipolar diffusion. We also studied different amplitudes and spectra of initially-supplied turbulence that is allowed to decay freely. Some of our key findings, including a comparison of clouds with strong and weak magnetic fields, are listed below.

\begin{itemize}

\item The column density PDFs for clouds with supercritical mass-to-flux ($\beta_{0} > 1$) ratio have a power law with indices $\alpha = 1.7$ and $\alpha = 2.2 $ for nonlinear and linear  turbulence, respectively. These power laws develop quickly both in time as well as in column density evolution, so that the PDFs are like a pure power law except at the lowest values.

\item Clouds with subcritical mass-to-flux ratio and linear perturbations (model V1) have a PDF that is a steep power law with index $\alpha = 4.3$. Similarly, for the subcritical model K1 with nonlinear white noise spectrum $v_{k}^{2} \propto k^{0}$, the PDF is also steep with index $\alpha = 4.0$. The steep slope of these subcritical models (V1 and K1) is indicative of the process that magnetic support restricts the rate of core and star formation in these clouds. This is consistent with the fact that clouds with steeper slopes ($\alpha= 3.9$ and $\alpha=3.0$), like Polaris and Pipe, respectively \citep{lom15}, have minimum star formation activity.

\item The fitting of the column density PDF of supercritical clouds or subcritical clouds with linear perturbations is best done by a modified lognormal power law (MLP) function \citep{bas15}. The MLP is a pure lognormal in one limit and pure power law in another, depending on the values of its three parameters. The value of the parameter $\alpha$ represents the slope of the power-law profile of all models studied here. The parameter $\alpha$ has a typical variation of $10\% - 20\%$ depending on the different realizations of the initial perturbation and fitting routines.
	

\item Only in the case of a subcritical (strong magnetic field) model V4 with nonlinear (supersonic) perturbations with turbulent spectrum $v_{k}^{2} \propto k^{-4}$ does the column density PDF retain a lognormal shape for the major part of its evolutionary phase as it oscillates due to the action of the magnetic field and turbulence. Eventually, the PDF develops a power-law tail due to gravitationally driven ambipolar diffusion, where the neutrals drift past the field lines and create subregions of supercritical pockets. In these subregions the PDF has an index $\alpha \simeq 2$, similar to supercritcal clouds.

\item A nonlinear velocity perturbation with turbulent spectrum $v_{k}^{2} \propto k^{-4}$ causes a much wider spread of the column density PDF in both supercritical and subcritical clouds compared to clouds with linear perturbations. Furthermore, the strength of the strong velocity perturbation shortens the core formation time \citep{kud11}.  


%

\end{itemize}

\section*{Acknowledgements}

We thank Deepakshi Madaan for fruitful discussions, and the anonymous referee for constructive comments. Computations were carried out using facilities of the Shared Hierarchical Academic Research Network (SHARCNET) and the Center for Computational Astrophysics, National Astronomical Observatory of Japan. SB is supported by a Discovery Grant from NSERC.



\bibliographystyle{mnras}
\bibliography{myrefs.bib} 

\begin{thebibliography}{}
\makeatletter
\relax
\def\mn@urlcharsother{\let\do\@makeother \do\$\do\&\do\#\do\^\do\_\do\%\do\~}
\def\mn@doi{\begingroup\mn@urlcharsother \@ifnextchar [ {\mn@doi@}
  {\mn@doi@[]}}
\def\mn@doi@[#1]#2{\def\@tempa{#1}\ifx\@tempa\@empty \href
  {http://dx.doi.org/#2} {doi:#2}\else \href {http://dx.doi.org/#2} {#1}\fi
  \endgroup}
\def\mn@eprint#1#2{\mn@eprint@#1:#2::\@nil}
\def\mn@eprint@arXiv#1{\href {http://arxiv.org/abs/#1} {{\tt arXiv:#1}}}
\def\mn@eprint@dblp#1{\href {http://dblp.uni-trier.de/rec/bibtex/#1.xml}
  {dblp:#1}}
\def\mn@eprint@#1:#2:#3:#4\@nil{\def\@tempa {#1}\def\@tempb {#2}\def\@tempc
  {#3}\ifx \@tempc \@empty \let \@tempc \@tempb \let \@tempb \@tempa \fi \ifx
  \@tempb \@empty \def\@tempb {arXiv}\fi \@ifundefined
  {mn@eprint@\@tempb}{\@tempb:\@tempc}{\expandafter \expandafter \csname
  mn@eprint@\@tempb\endcsname \expandafter{\@tempc}}}

\bibitem[\protect\citeauthoryear{{Alves}, {Lada}  \& {Lada}}{{Alves}
  et~al.}{2001}]{alv01}
{Alves} J.~F.,  {Lada} C.~J.,   {Lada} E.~A.,  2001, \nat, \href
  {http://adsabs.harvard.edu/abs/2001Natur.409..159A} {409, 159}

\bibitem[\protect\citeauthoryear{{Alves}, {Lombardi}  \& {Lada}}{{Alves}
  et~al.}{2014}]{alv14}
{Alves} J.,  {Lombardi} M.,   {Lada} C.~J.,  2014, \mn@doi [\aap]
  {10.1051/0004-6361/201322159}, \href
  {http://adsabs.harvard.edu/abs/2014A%26A...565A..18A} {565, A18}

\bibitem[\protect\citeauthoryear{{Alves}, {Lombardi}  \& {Lada}}{{Alves}
  et~al.}{2017}]{alv17}
{Alves} J.,  {Lombardi} M.,   {Lada} C.~J.,  2017, \mn@doi [\aap]
  {10.1051/0004-6361/201731436}, \href
  {http://adsabs.harvard.edu/abs/2017A%26A...606L...2A} {606, L2}

\bibitem[\protect\citeauthoryear{{Auddy}, {Basu}  \& {Kudoh}}{{Auddy}
  et~al.}{2016}]{aud16}
{Auddy} S.,  {Basu} S.,   {Kudoh} T.,  2016, \mn@doi [\apj]
  {10.3847/0004-637X/831/1/46}, \href
  {http://adsabs.harvard.edu/abs/2016ApJ...831...46A} {831, 46}

\bibitem[\protect\citeauthoryear{{Bacmann}, {Andr{\'e}}, {Puget}, {Abergel},
  {Bontemps}  \& {Ward-Thompson}}{{Bacmann} et~al.}{2000}]{bac00}
{Bacmann} A.,  {Andr{\'e}} P.,  {Puget} J.-L.,  {Abergel} A.,  {Bontemps} S.,
  {Ward-Thompson} D.,  2000, \aap, \href
  {http://adsabs.harvard.edu/abs/2000A%26A...361..555B} {361, 555}

\bibitem[\protect\citeauthoryear{{Ballesteros-Paredes}, {V{\'a}zquez-Semadeni},
  {Gazol}, {Hartmann}, {Heitsch}  \& {Col{\'{\i}}n}}{{Ballesteros-Paredes}
  et~al.}{2011}]{par11}
{Ballesteros-Paredes} J.,  {V{\'a}zquez-Semadeni} E.,  {Gazol} A.,  {Hartmann}
  L.~W.,  {Heitsch} F.,   {Col{\'{\i}}n} P.,  2011, \mn@doi [\mnras]
  {10.1111/j.1365-2966.2011.19141.x}, \href
  {http://adsabs.harvard.edu/abs/2011MNRAS.416.1436B} {416, 1436}

\bibitem[\protect\citeauthoryear{{Basu}}{{Basu}}{1997}]{bas97}
{Basu} S.,  1997, \mn@doi [\apj] {10.1086/304420}, \href
  {http://adsabs.harvard.edu/abs/1997ApJ...485..240B} {485, 240}

\bibitem[\protect\citeauthoryear{{Basu} \& {Mouschovias}}{{Basu} \&
  {Mouschovias}}{1994}]{bas94}
{Basu} S.,  {Mouschovias} T.~C.,  1994, \mn@doi [\apj] {10.1086/174611}, \href
  {http://adsabs.harvard.edu/abs/1994ApJ...432..720B} {432, 720}

\bibitem[\protect\citeauthoryear{{Basu} \& {Mouschovias}}{{Basu} \&
  {Mouschovias}}{1995}]{bas95}
{Basu} S.,  {Mouschovias} T.~C.,  1995, \mn@doi [\apj] {10.1086/176387}, \href
  {http://adsabs.harvard.edu/abs/1995ApJ...453..271B} {453, 271}

\bibitem[\protect\citeauthoryear{{Basu}, {Ciolek}, {Dapp}  \& {Wurster}}{{Basu}
  et~al.}{2009}]{bas09}
{Basu} S.,  {Ciolek} G.~E.,  {Dapp} W.~B.,   {Wurster} J.,  2009, \mn@doi [\na]
  {10.1016/j.newast.2009.01.004}, \href
  {http://adsabs.harvard.edu/abs/2009NewA...14..483B} {14, 483}

\bibitem[\protect\citeauthoryear{{Basu}, {Gil}  \& {Auddy}}{{Basu}
  et~al.}{2015}]{bas15}
{Basu} S.,  {Gil} M.,   {Auddy} S.,  2015, \mn@doi [\mnras]
  {10.1093/mnras/stv445}, \href
  {http://adsabs.harvard.edu/abs/2015MNRAS.449.2413B} {449, 2413}

\bibitem[\protect\citeauthoryear{{Bonnor}}{{Bonnor}}{1956}]{bon56}
{Bonnor} W.~B.,  1956, \mn@doi [\mnras] {10.1093/mnras/116.3.351}, \href
  {http://adsabs.harvard.edu/abs/1956MNRAS.116..351B} {116, 351}

\bibitem[\protect\citeauthoryear{{Ciolek} \& {Basu}}{{Ciolek} \&
  {Basu}}{2006}]{cio06}
{Ciolek} G.~E.,  {Basu} S.,  2006, \mn@doi [\apj] {10.1086/507865}, \href
  {http://adsabs.harvard.edu/abs/2006ApJ...652..442C} {652, 442}

\bibitem[\protect\citeauthoryear{{Clauset}, {Shalizi}  \& {Newman}}{{Clauset}
  et~al.}{2009}]{cla09}
{Clauset} A.,  {Shalizi} C.~R.,   {Newman} M.~E.~J.,  2009, \mn@doi [SIAM
  Review] {10.1137/070710111}, \href
  {http://adsabs.harvard.edu/abs/2009SIAMR..51..661C} {51, 661}

\bibitem[\protect\citeauthoryear{{Collins}, {Kritsuk}, {Padoan}, {Li}, {Xu},
  {Ustyugov}  \& {Norman}}{{Collins} et~al.}{2012}]{col12}
{Collins} D.~C.,  {Kritsuk} A.~G.,  {Padoan} P.,  {Li} H.,  {Xu} H.,
  {Ustyugov} S.~D.,   {Norman} M.~L.,  2012, \mn@doi [\apj]
  {10.1088/0004-637X/750/1/13}, \href
  {http://adsabs.harvard.edu/abs/2012ApJ...750...13C} {750, 13}

\bibitem[\protect\citeauthoryear{{Crutcher}}{{Crutcher}}{2012}]{cru12}
{Crutcher} R.~M.,  2012, \mn@doi [\araa] {10.1146/annurev-astro-081811-125514},
  \href {http://adsabs.harvard.edu/abs/2012ARA%26A..50...29C} {50, 29}

\bibitem[\protect\citeauthoryear{{Dapp} \& {Basu}}{{Dapp} \&
  {Basu}}{2009}]{dap09}
{Dapp} W.~B.,  {Basu} S.,  2009, \mn@doi [\mnras]
  {10.1111/j.1365-2966.2009.14616.x}, \href
  {http://adsabs.harvard.edu/abs/2009MNRAS.395.1092D} {395, 1092}

\bibitem[\protect\citeauthoryear{{Ebert}}{{Ebert}}{1955}]{ebe55}
{Ebert} R.,  1955, \zap, \href
  {http://adsabs.harvard.edu/abs/1955ZA.....37..217E} {37, 217}

\bibitem[\protect\citeauthoryear{{Federrath} \& {Klessen}}{{Federrath} \&
  {Klessen}}{2012}]{fed12}
{Federrath} C.,  {Klessen} R.~S.,  2012, \mn@doi [\apj]
  {10.1088/0004-637X/761/2/156}, \href
  {http://adsabs.harvard.edu/abs/2012ApJ...761..156F} {761, 156}

\bibitem[\protect\citeauthoryear{{Federrath} \& {Klessen}}{{Federrath} \&
  {Klessen}}{2013}]{fed13}
{Federrath} C.,  {Klessen} R.~S.,  2013, \mn@doi [\apj]
  {10.1088/0004-637X/763/1/51}, \href
  {http://adsabs.harvard.edu/abs/2013ApJ...763...51F} {763, 51}

\bibitem[\protect\citeauthoryear{{Federrath}, {Klessen}  \&
  {Schmidt}}{{Federrath} et~al.}{2008}]{fed08}
{Federrath} C.,  {Klessen} R.~S.,   {Schmidt} W.,  2008, \mn@doi [\apjl]
  {10.1086/595280}, \href {http://adsabs.harvard.edu/abs/2008ApJ...688L..79F}
  {688, L79}

\bibitem[\protect\citeauthoryear{{Fischera}}{{Fischera}}{2014}]{fis14a}
{Fischera} J.,  2014, \mn@doi [\aap] {10.1051/0004-6361/201321417}, \href
  {http://adsabs.harvard.edu/abs/2014A%26A...565A..24F} {565, A24}

\bibitem[\protect\citeauthoryear{{Hennebelle} \& {Chabrier}}{{Hennebelle} \&
  {Chabrier}}{2008}]{hen08}
{Hennebelle} P.,  {Chabrier} G.,  2008, \mn@doi [\apj] {10.1086/589916}, \href
  {http://adsabs.harvard.edu/abs/2008ApJ...684..395H} {684, 395}

\bibitem[\protect\citeauthoryear{{Hennebelle} \& {Chabrier}}{{Hennebelle} \&
  {Chabrier}}{2011}]{hen11}
{Hennebelle} P.,  {Chabrier} G.,  2011, \mn@doi [\apjl]
  {10.1088/2041-8205/743/2/L29}, \href
  {http://adsabs.harvard.edu/abs/2011ApJ...743L..29H} {743, L29}

\bibitem[\protect\citeauthoryear{{Hunter}}{{Hunter}}{1977}]{hun77}
{Hunter} C.,  1977, \mn@doi [\apj] {10.1086/155739}, \href
  {http://adsabs.harvard.edu/abs/1977ApJ...218..834H} {218, 834}

\bibitem[\protect\citeauthoryear{Johnson, Kotz  \& Balakrishnan}{Johnson
  et~al.}{2002}]{jon02}
Johnson N.~L.,  Kotz S.,   Balakrishnan N.,  2002, {Continuous multivariate
  distributions, volume 1, models and applications}.
 Vol. 59, New York: John Wiley \& Sons

\bibitem[\protect\citeauthoryear{{Kainulainen}, {Beuther}, {Henning}  \&
  {Plume}}{{Kainulainen} et~al.}{2009}]{kai09}
{Kainulainen} J.,  {Beuther} H.,  {Henning} T.,   {Plume} R.,  2009, \mn@doi
  [\aap] {10.1051/0004-6361/200913605}, \href
  {http://adsabs.harvard.edu/abs/2009A%26A...508L..35K} {508, L35}

\bibitem[\protect\citeauthoryear{{K{\"o}nyves} et~al.,}{{K{\"o}nyves}
  et~al.}{2015}]{kon15}
{K{\"o}nyves} V.,  et~al., 2015, \mn@doi [\aap] {10.1051/0004-6361/201525861},
  \href {http://adsabs.harvard.edu/abs/2015A%26A...584A..91K} {584, A91}

\bibitem[\protect\citeauthoryear{{Kritsuk}, {Norman}  \& {Wagner}}{{Kritsuk}
  et~al.}{2011}]{kri11}
{Kritsuk} A.~G.,  {Norman} M.~L.,   {Wagner} R.,  2011, \mn@doi [\apjl]
  {10.1088/2041-8205/727/1/L20}, \href
  {http://adsabs.harvard.edu/abs/2011ApJ...727L..20K} {727, L20}

\bibitem[\protect\citeauthoryear{{Kudoh} \& {Basu}}{{Kudoh} \&
  {Basu}}{2003}]{kud03}
{Kudoh} T.,  {Basu} S.,  2003, \mn@doi [\apj] {10.1086/377495}, \href
  {http://adsabs.harvard.edu/abs/2003ApJ...595..842K} {595, 842}

\bibitem[\protect\citeauthoryear{{Kudoh} \& {Basu}}{{Kudoh} \&
  {Basu}}{2006}]{kud06}
{Kudoh} T.,  {Basu} S.,  2006, \mn@doi [\apj] {10.1086/500726}, \href
  {http://adsabs.harvard.edu/abs/2006ApJ...642..270K} {642, 270}

\bibitem[\protect\citeauthoryear{{Kudoh} \& {Basu}}{{Kudoh} \&
  {Basu}}{2008}]{kud08}
{Kudoh} T.,  {Basu} S.,  2008, \mn@doi [\apjl] {10.1086/589618}, \href
  {http://adsabs.harvard.edu/abs/2008ApJ...679L..97K} {679, L97}

\bibitem[\protect\citeauthoryear{{Kudoh} \& {Basu}}{{Kudoh} \&
  {Basu}}{2011}]{kud11}
{Kudoh} T.,  {Basu} S.,  2011, \mn@doi [\apj] {10.1088/0004-637X/728/2/123},
  \href {http://adsabs.harvard.edu/abs/2011ApJ...728..123K} {728, 123}

\bibitem[\protect\citeauthoryear{{Kudoh} \& {Basu}}{{Kudoh} \&
  {Basu}}{2014}]{kud14}
{Kudoh} T.,  {Basu} S.,  2014, \mn@doi [\apj] {10.1088/0004-637X/794/2/127},
  \href {http://adsabs.harvard.edu/abs/2014ApJ...794..127K} {794, 127}

\bibitem[\protect\citeauthoryear{{Kudoh}, {Basu}, {Ogata}  \& {Yabe}}{{Kudoh}
  et~al.}{2007}]{kud07}
{Kudoh} T.,  {Basu} S.,  {Ogata} Y.,   {Yabe} T.,  2007, \mn@doi [\mnras]
  {10.1111/j.1365-2966.2007.12119.x}, \href
  {http://adsabs.harvard.edu/abs/2007MNRAS.380..499K} {380, 499}

\bibitem[\protect\citeauthoryear{{Langer}}{{Langer}}{1978}]{lan78}
{Langer} W.~D.,  1978, \mn@doi [\apj] {10.1086/156471}, \href
  {http://adsabs.harvard.edu/abs/1978ApJ...225...95L} {225, 95}

\bibitem[\protect\citeauthoryear{{Larson}}{{Larson}}{1969}]{lar69}
{Larson} R.~B.,  1969, \mn@doi [\mnras] {10.1093/mnras/145.3.271}, \href
  {http://adsabs.harvard.edu/abs/1969MNRAS.145..271L} {145, 271}

\bibitem[\protect\citeauthoryear{{Lombardi}, {Bouy}, {Alves}  \&
  {Lada}}{{Lombardi} et~al.}{2014}]{lom14}
{Lombardi} M.,  {Bouy} H.,  {Alves} J.,   {Lada} C.~J.,  2014, \mn@doi [\aap]
  {10.1051/0004-6361/201323293}, \href
  {http://adsabs.harvard.edu/abs/2014A%26A...566A..45L} {566, A45}

\bibitem[\protect\citeauthoryear{{Lombardi}, {Alves}  \& {Lada}}{{Lombardi}
  et~al.}{2015}]{lom15}
{Lombardi} M.,  {Alves} J.,   {Lada} C.~J.,  2015, \mn@doi [\aap]
  {10.1051/0004-6361/201525650}, \href
  {http://adsabs.harvard.edu/abs/2015A%26A...576L...1L} {576, L1}

\bibitem[\protect\citeauthoryear{{Mestel} \& {Spitzer}}{{Mestel} \&
  {Spitzer}}{1956}]{mes56}
{Mestel} L.,  {Spitzer} Jr. L.,  1956, \mn@doi [\mnras]
  {10.1093/mnras/116.5.503}, \href
  {http://adsabs.harvard.edu/abs/1956MNRAS.116..503M} {116, 503}

\bibitem[\protect\citeauthoryear{{Molina}, {Glover}, {Federrath}  \&
  {Klessen}}{{Molina} et~al.}{2012}]{mol12}
{Molina} F.~Z.,  {Glover} S.~C.~O.,  {Federrath} C.,   {Klessen} R.~S.,  2012,
  \mn@doi [\mnras] {10.1111/j.1365-2966.2012.21075.x}, \href
  {http://adsabs.harvard.edu/abs/2012MNRAS.423.2680M} {423, 2680}

\bibitem[\protect\citeauthoryear{{Mouschovias} \& {Spitzer}}{{Mouschovias} \&
  {Spitzer}}{1976}]{mou76}
{Mouschovias} T.~C.,  {Spitzer} Jr. L.,  1976, \mn@doi [\apj] {10.1086/154835},
  \href {http://adsabs.harvard.edu/abs/1976ApJ...210..326M} {210, 326}

\bibitem[\protect\citeauthoryear{{Mouschovias}, {Ciolek}  \&
  {Morton}}{{Mouschovias} et~al.}{2011}]{mou11}
{Mouschovias} T.~C.,  {Ciolek} G.~E.,   {Morton} S.~A.,  2011, \mn@doi [\mnras]
  {10.1111/j.1365-2966.2011.18817.x}, \href
  {http://adsabs.harvard.edu/abs/2011MNRAS.415.1751M} {415, 1751}

\bibitem[\protect\citeauthoryear{{Nakamura} \& {Li}}{{Nakamura} \&
  {Li}}{2005}]{nak05}
{Nakamura} F.,  {Li} Z.-Y.,  2005, \mn@doi [\apj] {10.1086/432606}, \href
  {http://adsabs.harvard.edu/abs/2005ApJ...631..411N} {631, 411}

\bibitem[\protect\citeauthoryear{{Nakano} \& {Nakamura}}{{Nakano} \&
  {Nakamura}}{1978}]{nak78}
{Nakano} T.,  {Nakamura} T.,  1978, \pasj, \href
  {http://adsabs.harvard.edu/abs/1978PASJ...30..671N} {30, 671}

\bibitem[\protect\citeauthoryear{{Padoan} \& {Nordlund}}{{Padoan} \&
  {Nordlund}}{1999}]{pad99}
{Padoan} P.,  {Nordlund} {\AA}.,  1999, \mn@doi [\apj] {10.1086/307956}, \href
  {http://adsabs.harvard.edu/abs/1999ApJ...526..279P} {526, 279}

\bibitem[\protect\citeauthoryear{{Padoan} \& {Nordlund}}{{Padoan} \&
  {Nordlund}}{2002}]{pad02}
{Padoan} P.,  {Nordlund} {\AA}.,  2002, \mn@doi [\apj] {10.1086/341790}, \href
  {http://adsabs.harvard.edu/abs/2002ApJ...576..870P} {576, 870}

\bibitem[\protect\citeauthoryear{{Padoan} \& {Nordlund}}{{Padoan} \&
  {Nordlund}}{2011}]{pad11}
{Padoan} P.,  {Nordlund} {\AA}.,  2011, \mn@doi [\apj]
  {10.1088/0004-637X/730/1/40}, \href
  {http://adsabs.harvard.edu/abs/2011ApJ...730...40P} {730, 40}

\bibitem[\protect\citeauthoryear{{Padoan}, {Jones}  \& {Nordlund}}{{Padoan}
  et~al.}{1997}]{pad97}
{Padoan} P.,  {Jones} B.~J.~T.,   {Nordlund} {\AA}.~P.,  1997, \mn@doi [\apj]
  {10.1086/303482}, \href {http://adsabs.harvard.edu/abs/1997ApJ...474..730P}
  {474, 730}

\bibitem[\protect\citeauthoryear{{Passot} \& {V{\'a}zquez-Semadeni}}{{Passot}
  \& {V{\'a}zquez-Semadeni}}{1998}]{pas98}
{Passot} T.,  {V{\'a}zquez-Semadeni} E.,  1998, \mn@doi [\pre]
  {10.1103/PhysRevE.58.4501}, \href
  {http://adsabs.harvard.edu/abs/1998PhRvE..58.4501P} {58, 4501}

\bibitem[\protect\citeauthoryear{{Pokhrel} et~al.,}{{Pokhrel}
  et~al.}{2016}]{pok16}
{Pokhrel} R.,  et~al., 2016, \mn@doi [\mnras] {10.1093/mnras/stw1303}, \href
  {http://adsabs.harvard.edu/abs/2016MNRAS.461...22P} {461, 22}

\bibitem[\protect\citeauthoryear{{Scalo}, {V{\'a}zquez-Semadeni}, {Chappell}
  \& {Passot}}{{Scalo} et~al.}{1998}]{sca98}
{Scalo} J.,  {V{\'a}zquez-Semadeni} E.,  {Chappell} D.,   {Passot} T.,  1998,
  \mn@doi [\apj] {10.1086/306099}, \href
  {http://adsabs.harvard.edu/abs/1998ApJ...504..835S} {504, 835}

\bibitem[\protect\citeauthoryear{{Shu}}{{Shu}}{1977}]{shu77}
{Shu} F.~H.,  1977, \mn@doi [\apj] {10.1086/155274}, \href
  {http://adsabs.harvard.edu/abs/1977ApJ...214..488S} {214, 488}

\bibitem[\protect\citeauthoryear{{Spitzer}}{{Spitzer}}{1942}]{spi42}
{Spitzer} Jr. L.,  1942, \mn@doi [\apj] {10.1086/144407}, \href
  {http://adsabs.harvard.edu/abs/1942ApJ....95..329S} {95, 329}

\bibitem[\protect\citeauthoryear{{Strittmatter}}{{Strittmatter}}{1966}]{str66}
{Strittmatter} P.~A.,  1966, \mn@doi [\mnras] {10.1093/mnras/132.2.359}, \href
  {http://adsabs.harvard.edu/abs/1966MNRAS.132..359S} {132, 359}

\bibitem[\protect\citeauthoryear{{Tafalla}, {Myers}, {Caselli}, {Walmsley}  \&
  {Comito}}{{Tafalla} et~al.}{2002}]{taf02}
{Tafalla} M.,  {Myers} P.~C.,  {Caselli} P.,  {Walmsley} C.~M.,   {Comito} C.,
  2002, \mn@doi [\apj] {10.1086/339321}, \href
  {http://adsabs.harvard.edu/abs/2002ApJ...569..815T} {569, 815}

\bibitem[\protect\citeauthoryear{{Tassis}, {Christie}, {Urban}, {Pineda},
  {Mouschovias}, {Yorke}  \& {Martel}}{{Tassis} et~al.}{2010}]{tas10}
{Tassis} K.,  {Christie} D.~A.,  {Urban} A.,  {Pineda} J.~L.,  {Mouschovias}
  T.~C.,  {Yorke} H.~W.,   {Martel} H.,  2010, \mn@doi [\mnras]
  {10.1111/j.1365-2966.2010.17181.x}, \href
  {http://adsabs.harvard.edu/abs/2010MNRAS.408.1089T} {408, 1089}

\bibitem[\protect\citeauthoryear{{Tomisaka}, {Ikeuchi}  \&
  {Nakamura}}{{Tomisaka} et~al.}{1988}]{tom88}
{Tomisaka} K.,  {Ikeuchi} S.,   {Nakamura} T.,  1988, \mn@doi [\apj]
  {10.1086/166923}, \href {http://adsabs.harvard.edu/abs/1988ApJ...335..239T}
  {335, 239}

\bibitem[\protect\citeauthoryear{{Vazquez-Semadeni}}{{Vazquez-Semadeni}}{1994}]{sem94}
{Vazquez-Semadeni} E.,  1994, \mn@doi [\apj] {10.1086/173847}, \href
  {http://adsabs.harvard.edu/abs/1994ApJ...423..681V} {423, 681}

\bibitem[\protect\citeauthoryear{{Ward-Thompson}, {Motte}  \&
  {Andre}}{{Ward-Thompson} et~al.}{1999}]{war99}
{Ward-Thompson} D.,  {Motte} F.,   {Andre} P.,  1999, \mn@doi [\mnras]
  {10.1046/j.1365-8711.1999.02412.x}, \href
  {http://adsabs.harvard.edu/abs/1999MNRAS.305..143W} {305, 143}

\bibitem[\protect\citeauthoryear{{Ward}, {Wadsley}  \& {Sills}}{{Ward}
  et~al.}{2014}]{war14}
{Ward} R.~L.,  {Wadsley} J.,   {Sills} A.,  2014, \mn@doi [\mnras]
  {10.1093/mnras/stu1868}, \href
  {http://adsabs.harvard.edu/abs/2014MNRAS.445.1575W} {445, 1575}

\makeatother
\end{thebibliography}




\appendix

\section{Analytic Model of the PDF}

Molecular clouds contain many cores that can be approximately described as isothermal spheres. There are several analytic column density profiles to fit prestellar cores. One model to fit such cores is the Bonnor-Ebert sphere \citep{ebe55,bon56}. This model assumes isothermal gas spheres bounded by external pressure and a hydrostatic equilibrium of gravity and thermal pressure. \cite{dap09} proposed a new three-parameter analytic formula to fit the column density profiles of prestellar cores. This model does not assume the cloud to be in equilibrium and can fit the dynamical states of the nonequilibrium collapse solutions \citep{lar69} as well.
\cite{fis14a} discusses the PDF of the mass surface density of molecular clouds. The column density PDF of the molecular clouds show two distinct features, involving a broad distribution around the peak and a power-law tail at the high end. The first aspect can be attributed to the turbulence of the cloud, while the tail develops because of the gravitationally condensed structures. These condensed structures are modelled as spheres or cylinders with a truncated radial density. \citet{fis14a} provided an analytic model of the PDF of the condensed structures, by either considering them as spheres or cylinders with a truncated radial profile. He concluded that the asymptotic behaviour  of the logarithmic PDF ($\Sigma P$($\Sigma$)) in the limit of high column density has a power-law index $\alpha = $($p+1$)/($p-1$) for spheres or $\alpha = p$/($p-1$)$ - 1$ for cylinders, 
where $p$ is the power-law index of the column density profile of the condensed structures.
 
For clouds with a preferred direction of magnetic field the cloud is flattened along the direction of the magnetic field and settles in to a hydrostatic equilibrium 
\citep{dap09}. In our simulation, the magnetic field is oriented along the $z-$axis, resulting in a collapsing core resembling a disk as it is flattened along the magnetic field.

\subsection{ Disk Geometry}\label{disk geometry}
 The generic face-on column density profile for disk shaped flattened core is 

\begin{equation}\label{disk}
\begin{split}
 \Sigma\left(r\right) & = \frac{\Sigma_c }{\sqrt{\left(1+\left(\frac{r}{a}\right)^{p}\right)}}\ & r \leq R, \\ 
		   & =  0, \,	& r > R,
\end{split}
\end{equation}
where $p$ is the power-law index for the column density profile. The gravitational contraction of the cores lead to the formation of supercritical pockets of near uniform column density regions surrounded by a power-law profile. The index $p$ achieves the value $2$ just outside the uniform region $a$, but transitions to $\sim 1.5$ at the boundary of the supercritical cores \citep{bas95,bas97}. 
Assuming vertical hydrostatic equilibrium, the volume density is proportional to the square of the column density \citep{spi42} so that 

\begin{equation}
c_{s}^{2} \rho = \frac{\pi}{2}G \Sigma^{2},
\end{equation} 
and the corresponding density is given as 

\begin{equation}
	\rho\left(r\right) = \frac{\pi G}{2 c_{s}^{2}} \frac{\Sigma_{c}^{2}}{1+\left(\frac{r}{a}\right)^{p}}.
\end{equation}

One can model the molecular cloud as an ensemble of condensed disks. The column density PDF is then 

 \begin{equation}\label{pdfdisk}
P \left(\Sigma\right) = \frac{d N^{'}}{d \Sigma} =-P$($r$)$\left(\frac{d\Sigma}{dr}\right)^{-1},
\end{equation}
where $d N^{\prime} = d N $/$ N_{\rm{total}}$. Let $r$/$a$ be the normalized impact radius where $a=kc_s$/($\sqrt{ G \rho_c}$) is the size of the flat region. $P$($r$)$ dr$ is the probability to measure the column density at a impact radius of $r$. 
Using equation~\ref{disk} we get 

\begin{equation}\label{diffdisk}
\frac{d \Sigma}{d r} = -\frac{\Sigma_c p \left(\frac{r}{a}\right)^{p-1}}{2 a \left(1+\left(\frac{r}{a}\right)^p\right)^{\frac{3}{2}}}.
\end{equation} 

On substituting Equation \ref{diffdisk} in Equation \ref{pdfdisk} and using $P$($r$)$ = 2 \pi r $/($\pi R^2$) we establish the PDF as


\begin{equation}\label{diskpdf}
P$($\Sigma$)$ = \frac{4}{c^2 \Sigma_{c}p} \left[\left( \frac{\Sigma_c}{\Sigma}\right)^{2}-1\right]^{\left(\frac{2}{p-1}\right)} \left(\frac{\Sigma_c}{\Sigma}\right)^3.
\end{equation} 
The power-law index for the column density profile (Equation \ref{disk}) $p=2$ and $p=1$ corresponds to just outside the uniform region and the boundary of the supercritical core respectively. Depending on the region of interest we can estimate the power-law index the column density PDF (using Equation \ref{diskpdf}). For $p=2$ we get

\begin{equation}\label{n2diskpdf}
P\left(\Sigma\right)_{p=2} =  \frac{2}{c^2 }\frac{\Sigma_c^{2}}{\Sigma^{3}}.
\end{equation}  
In the regions just outside the uniform region, $P $($\Sigma$)$ \propto \Sigma ^{-3}$.
For $p=1$ we find the column density PDF as 

\begin{equation}\label{n1diskpdf}
P$($\Sigma$)$_{p=1} =  \frac{4\Sigma_{c}^{2}}{c^2} \frac{\left(\Sigma_{c}^{2}-{\Sigma^{2}}\right)}{\Sigma^{5}}.
\end{equation}   

For $\Sigma\ll \Sigma_c$, i.e. corresponding to region well outside the center and closer to the core boundary, $P$($\Sigma$)$ \propto \Sigma^{-5}$.
\bsp	
\label{lastpage}
\end{document}